\documentclass[prl,reprint, amsmath,amssymb]{revtex4-1}
\usepackage{graphicx}
\usepackage{dcolumn}
\usepackage{bm}
\usepackage{mathrsfs}
\usepackage{hyperref}
\usepackage{epstopdf}
\allowdisplaybreaks[4]

\begin{document}
\title{Nonlinear emergent elasticity and structural transitions of skyrmion crystal under uniaxial distortion}
\author{Yangfan Hu}
 \email[Corresponding author. ]{huyf3@mail.sysu.edu.cn}
\author{Xiaoming Lan}
\author{Biao Wang}
 \email[Corresponding author. ]{wangbiao@mail.sysu.edu.cn}
\affiliation{Sino-French Institute of Nuclear Engineering and Technology, Sun Yat-sen University, 519082, Zhuhai, China}

\begin{abstract}
    Emergent crystals are periodic alignment of ``emergent particles'', i.e., localized collective behavior of atoms or their charges/spins/orbits. These novel states of matter, widely observed in various systems, may deform under mechanical forces with elasticity strikingly different from that of the underlying material. However, their nonlinear and critical behaviors under strong fields are hitherto unclear. Here we theoretically study the nonlinear elasticity and structural transitions of skyrmion crystals (SkX) suffering uniaxial distortion by using three different methods. Under moderate tension, SkX behaves like a ductile material, with a negative crossover elastic stiffness and a negative emergent Poisson's ratio at appropriate conditions of magnetic field. Under strong straining, we observe at most six phase transitions, leading to appearance of four novel emergent crystals that are thermodynamically metastable. When subject to external loads, emergent crystals rotate globally, and their composing ``particles'' have unlimited deformability, which render their exotic polymorphism.
\end{abstract}
\maketitle
\noindent{\textbf{Introduction}}

A recent outburst of interest in emergent crystals occurs due to the discovery of skyrmions with various chirality\cite{muhlbauer2009skyrmion,kezsmarki2015neel}, commensurability\cite{von2015influence,langner2014coupled} and dimensionality\cite{muhlbauer2009skyrmion,tanigaki2015real}, and ``relatives'' of skyrmions\cite{lee2015zero,huang2017stabilization,ackerman2015self} in different systems\cite{bauerle1996laboratory,al2001skyrmions,rossler2006spontaneous,fu2016persistent,nych2017spontaneous}, especially in bulk\cite{muhlbauer2009skyrmion,tanigaki2015real,karube2016robust,johnson2013mnsb,leonov2015multiply}and geometrically confined\cite{yu2010real,heinze2011spontaneous,yu2011near,jiang2017skyrmions,du2015electrical} magnetic materials where they appear as localized spin textures with nontrivial topology. These topological objects behave as elementary particles in an emerging world, such that they form novel crystalline states at low temperature, and melt under heating\cite{ambrose2013melting}. On one side, these emergent crystals benefit from the exotic local properties of their composing particles (e.g., topological protection, mobility to electric current\cite{jonietz2010spin,zang2011dynamics}, and skyrmion Hall effects\cite{neubauer2009topological,litzius2017skyrmion,jiang2017direct}). On the other side, they give rise to novel macroscopic emergent properties when interacting with external fields\cite{shiomi2013topological,white2014electric,shibata2015large,okamura2016transition}. Systematic study of the deformation and instability of emergent crystals under different types of external fields provides a major challenge for us to understand the fundamental similarities and discrepancies between emergent crystals and ordinary crystals, and to develop reliable approaches toward effective manipulation of these new states of matter.

Skyrmion crystals (SkX) in chiral magnets and their layered structures are profoundly affected by deformation of the material. It is known that uniaxial loading\cite{nii2015uniaxial,chacon2015uniaxial,wang2018uniaxial}, as well as misfit strains in thin film structures\cite{karhu2010structure,huang2012extended} both significantly change the stability of SkX in the temperature-magnetic field phase diagram. Moreover, mechanical loads affect the generation\cite{liu2017chopping} and chirality\cite{chen2015unlocking} of skyrmion, and the elementary excitations\cite{zhang2017ultrasonic} in the SkX. A fundamental type of phenomena that connects or even explains the ones mentioned above is the ability of SkX to deform under strains: when subject to uniaxial tension, the SkX inside FeGe thin film is found to undergo significant deformation about 66 times larger than the deformation of the underlying atomic lattice\cite{shibata2015large}, and this emergent elasticity of SkX has also been observed in MnSi\cite{shibata2015large,fobes2017versatile,kang2017elastic}. At present stage, most of existing theories on this anomalous emergent deformation of SkX are restricted to its elastic range\cite{hu2016emergent,petrova2011spin}, while the nonlinear deformation and instability of SkX suffering moderate or strong elastic fields have never been elucidated.

In this work, we study the nonlinear elasticity and structural phase transitions of SkX in bulk helimagnets within an analytical theoretical framework, where the deformation of SkX is described by the emergent elastic strains and the emergent rotation angles\cite{hu2016emergent}. Developed upon a thermodynamic potential incorporating a comprehensive magnetoelastic functional for B20 helimagnets\cite{hu2017unified}, the framework has proved its effectiveness by quantitatively reproducing for MnSi the phase diagram\cite{muhlbauer2009skyrmion}, the variation of elastic stiffness\cite{nii2014elastic}, and the changed stability of SkX in thin films\cite{li2013robust,hu2018effect}. For SkX in MnSi suffering uniaxial tension, we find that the stress-emergent strain curve before phase transition resembles the stress-strain curve of typical ductile materials (e.g., aluminum alloy). However, in most of the conditions studied, SkX possesses a negative emergent Poisson¡¯s ratio. And at high magnetic field and small uniaxial strain, it has an exotic negative crossover elastic stiffness. As the strain varies, the SkX undergoes five or six times of subsequent structural phase transitions, leading to appearance of four distorted SkX phases. Further analysis shows that these new states are thermodynamically metastable. We quantitatively reproduce the exotic emgergent elasticity of SkX observed in FeGe thin film suffering uniaxial tension\cite{shibata2015large}, and find that it is caused by a triangle-square phase transition.
\\
\\
\noindent{\textbf{Model}}

Consider a cubic helimagnet stabilized in the SkX phase suffering uniaxial distortion along the x-axis, the equilibrium state of the system is determined by the following rescaled Helmholtz free energy density functional (see the Methods section for derivation)\cite{hu2017unified}
\begin{equation}
    \begin{aligned}
        \widetilde w(\mathbf m)=&\widetilde w_0+\sum^3_{i=1}\left(\frac{\partial \mathbf m}{\partial r_i}\right)^2 +2\mathbf m\cdot (\mathbf\nabla\times \mathbf m)-2\mathbf b \cdot \mathbf m
        \\ &+t\mathbf m^2 +\mathbf m^4+\sum^3_{i=1}\left[\widetilde A_c m^4_i+\widetilde A_e \left(\frac{\partial m_i}{\partial r_i}\right)^2 \right]\\ &+\varepsilon_{11} [\widetilde K \mathbf m^2+\widetilde L_1 m^2_1+\widetilde L_2 m^2_3+\widetilde L_{O1} (m_{1,2}m_3
        \\ &-m_{1,3}m_2)+\widetilde L_{O2} (m_{3,1}m_2-m_{2,1}m_3)\\ &+\widetilde L_{O3} m_1(m_{2,3}-m_{3,2})],
    \end{aligned}
    \label{1}
\end{equation}
where $\mathbf m$ is the rescaled magnetization, $\mathbf\varepsilon_{11}$ is the strain component describing the uniaxial distortion, $t$ is the rescaled temperature, $\mathbf b$ is the rescaled magnetic field, and $\widetilde w_0$ denotes the part of free energy density independent of the magnetization. To describe the deformed SkX, $\mathbf m$ is expanded within the $n^{\text{th}}$ order Fourier representation as\cite{hu2016emergent,hu2018wave}

\begin{equation}
    \begin{aligned}
        \mathbf m=\mathbf m_0+\sum^n_{i=1}\sum^{n_i}_{j=1}\mathbf m_{\mathbf q_{ij}}e^{{\rm i} \mathbf q_{ij}(\bm \varepsilon^{ea})\cdot \mathbf r},
    \end{aligned}
    \label{2}
\end{equation}
or alternatively $\mathbf m=\mathbf(\mathbf m^a, \bm\varepsilon^{ea})$, where the vector $\mathbf m^a$ contains all components of the vectors of Fourier magnitude $\mathbf m_0$ and $\mathbf m_{\mathbf q_{ij}}$ for all $i$ and $j$, and the vector $\bm \varepsilon^{ea}$ contains all components of emergent elastic strains and the emergent rotation angles. For 2-D SkX. the exact form of $\mathbf m^a$ and $\bm \varepsilon^{ea}$ for $1^{st}$ order Fourier representation can be written as (see the Methods section for derivation)
\begin{equation}
    \begin{aligned}
        \bm \varepsilon^{ea}=[\varepsilon^e_{11},\varepsilon^e_{22},\varepsilon^e_{12},\omega^e]^T,
    \end{aligned}
    \label{3}
\end{equation}
\begin{equation}
    \begin{aligned}
        \mathbf m^a=[&m_{01},m_{02},m_{03},\tilde c^{re}_{111},\tilde c^{im}_{112},\tilde c^{re}_{113},
        \\ &\tilde c^{re}_{121},\tilde c^{im}_{122},\tilde c^{re}_{123},\tilde c^{re}_{131},\tilde c^{im}_{132},\tilde c^{re}_{133}]^T.
    \end{aligned}
    \label{4}
\end{equation}
One should keep in mind that the length of $\mathbf m^a$ depends on the order of Fourier representation, For $2^{nd}$, $3^{rd}$ and $4^{th}$ order Fourier representation, the number of components of $\mathbf m^a$ becomes 21, 30 and 48, respectively.
At given values of $t$, $\mathbf b$, and $\varepsilon_{11}$, the equilibrium state of the SkX phase is determined by minimizing $\bar{w} (\mathbf m^a,\bm \varepsilon^{ea})$ with respect to all components of $\mathbf m^a$ and $\bm \varepsilon^{ea}$, where $\bar{w} (\mathbf m^a,\bm \varepsilon^{ea})=\frac{1}{V}\int_{V}\widetilde{w} (\mathbf m^a,\bm \varepsilon^{ea})$ denotes the averaged free energy density of the system. To determine the thermodynamic equilibrium state of the system at given values of $t$, $\mathbf b$, and $\varepsilon_{11}$, we have to compare the minimized value $\bar{w}_{min} (\mathbf m^a,\bm \varepsilon^{ea})$ of the SkX phase with that of all the other possible magnetic phases such as the ferromagnetic phase and the conical phase. Concerning the large number of variables included in $\mathbf m^a$ and $\bm \varepsilon^{ea}$, the minimization problem is solved numerically. To guarantee that the solution obtained actually corresponds to a local minimum of the free energy of the system, the method of soft-mode analysis\cite{hu2018wave} is incorporated in the minimization process.
\newline
\newline
\noindent{\textbf{Nonlinear emergent elasticity of skyrmion crystal in MnSi}}

We analyze the nonlinear emergent elasticity and instability of SkX in a prototype helimagnet MnSi (thermodynamic parameters shown in ref.\cite{hu2017unified}) by calculating variation of the equilibrium field configuration of SkX (i.e., the values of components of $\bm \varepsilon^{ea}$ and $\mathbf m^a$) with the uniaxial distortion. We assume that the SkX is distributed in the x-y plane and the rescaled magnetic field satisfies $\mathbf b=[0, 0, b]^T$.

We first use three different methods to calculate the emergent elastic property of SkX at small uniaxial tensile strain, including the theory of emergent elasticity, the $4^{th}$ order Fourier representation based free energy minimization, and the Monte Carlo simulation based free energy minimization. As shown by the $\varepsilon_{11}-\varepsilon^e_{11}$ curves obtained at $t=0.5$ in FIG. 1(a), the results obtained from the three methods agree quantitatively well in the range $0\leq\varepsilon_{11}\leq 0.0016$, and are exactly equal to each other when $\varepsilon_{11}\to 0$. In this range, $\varepsilon^e_{11}$ is almost linear with $\varepsilon_{11}$ at all calculated conditions of $b$ and $\widetilde A_e$ according to the two free energy minimization methods. In a wider range of $\varepsilon_{11}$ (FIG. 1(a, b)), one discovers that the $\varepsilon_{11}-\varepsilon^e_{11}$ curves gradually presents nonlinearity, and the discrepancy between results obtained by the two methods gradually increases while maintaining qualitative agreement. This qualitative agreement repeats itself in the $\varepsilon_{11}-\varepsilon^e_{22}$ curves (FIG. 2(a)), demonstrating the effectiveness of the Fourier representation based free energy minimization method.

In the theory of emergent elasticity \cite{hu2016emergent}, the emergent elastic property of SkX is characterized by the crossover strain ratio (CSR) matrix, whose components are the linear coefficients relating $\bm \varepsilon^{ea}$ and the elastic strains of the underlying material. FIG. 1(a, c) shows that the strain-free CSR component $\lambda_{11}$ (inverse of the gradient of the $\varepsilon_{11}-\varepsilon^e_{11}$ curves at $\varepsilon_{11}=0$) decreases sharply with $b$, and is slightly affected by a change of exchange anisotropy (from $\widetilde A_e=0$ to $\widetilde A_e=-0.05$). Specifically, the $\varepsilon_{11}-\varepsilon^e_{11}$ curves calculated at $b=0.3$ correspond to a negative strain-free $\lambda_{11}$, indicating a compressing SkX in the x-axis when the underlying material is suffering elongation in the same direction.

\begin{figure}
    \centering
    \includegraphics[scale=0.28]{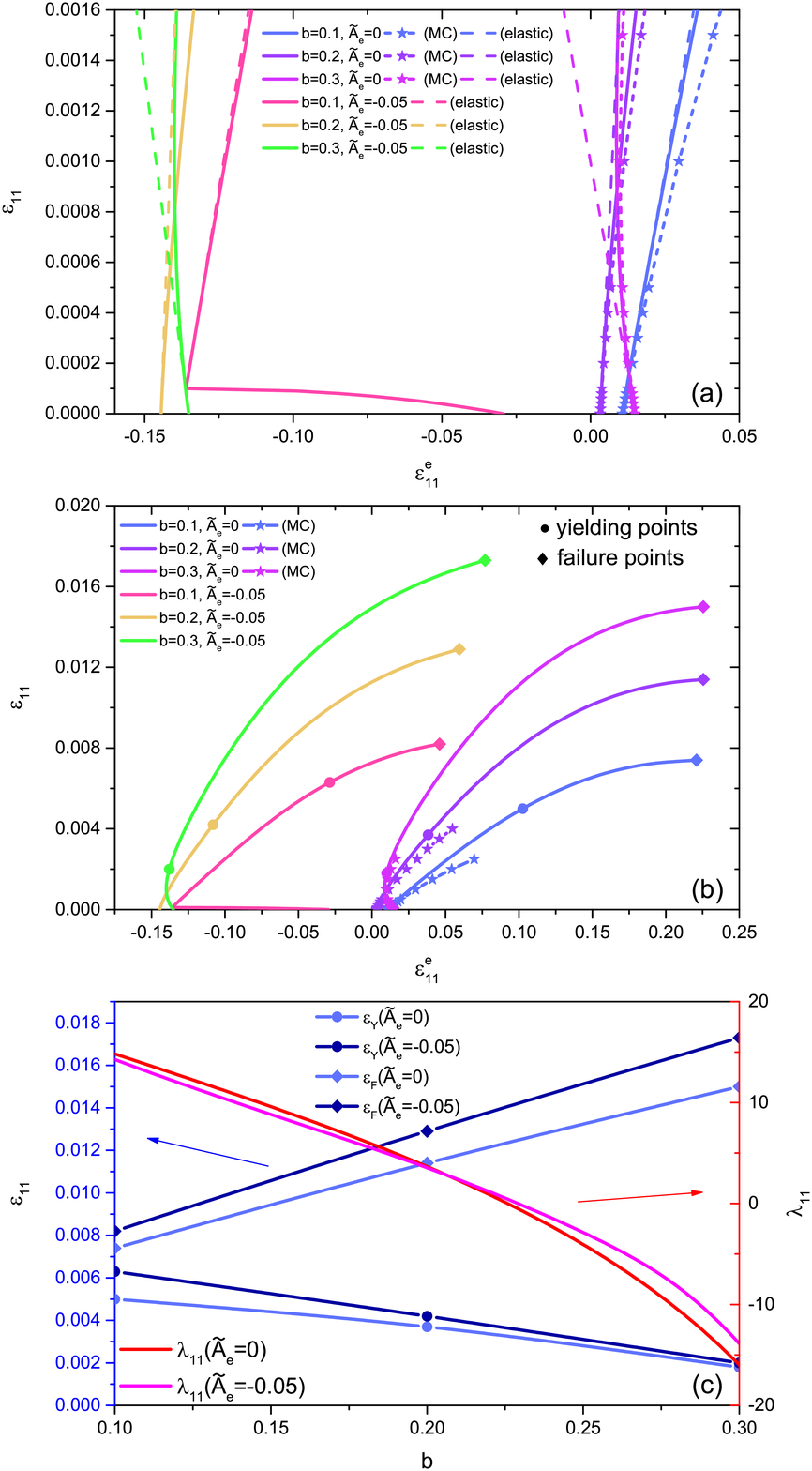}
    \caption{Variation of $\varepsilon^e_{11}$ with $\varepsilon_{11}$ and related emergent elastic properties for SkX in MnSi calculated at different values of rescaled magnetic field $b$ and rescaled exchange anisotropy coefficient $\widetilde A_e$. $(a)$ $\varepsilon_{11}-\varepsilon^e_{11}$ curve in the elastic stage, where solid curves are obtained through analytical free energy minimization, short-dashed curves with pentagram are obtained through free energy minimization with Monte Carlo simulation, and dashed lines are straight lines plotted with the CSR component $\lambda_{11}$ calculated at strain-free conditions. $(b)$ $\varepsilon_{11}-\varepsilon^e_{11}$ curve in the whole range of $\varepsilon_{11}$ before failure, where circular points mark the yield points defined by the intersection of the $\varepsilon_{11}-\varepsilon^e_{11}$ curve and the straight lines plotted with the strain-free CSR component $\lambda_{11}$ right shifted by $0.02$, and square points mark the failure points defined by the position where a structural phase transition occurs. $(c)$ Variation of the yielding strain $\varepsilon_Y$ and the failure strain $\varepsilon_F$ to the left axis, and the strain-free CSR component $\lambda_{11}$ to the right axis, with $b$.}
    \label{f1}
\end{figure}
In a wider range $0\leq\varepsilon_{11}\leq 0.02$, the $\varepsilon_{11}-\varepsilon^e_{11}$ curves (FIG. 1(b)) exhibit a similar shape to the stress-strain curve of any typical ductile metal (e.g., Aluminum alloy). Multiplying the vertical axis of FIG. 1(b) by the elastic constant $C_{11}$, we obtain the stress-emergent elastic strain curves for SkX under uniaxial distortion. At all calculated conditions of $b$ and $\widetilde A_e$, the emergent deformation of SkX with $\varepsilon_{11}$ can be divided in the three stages: linear elastic region, nonlinear elastic region, and ``failure'', where the strain corresponding to the point separating the linear elastic region and nonlinear elastic region is called emergent yielding strain $\varepsilon_Y$, and the strain corresponding to the failure point is called emergent failure strain $\varepsilon_F$. Imitating the definition of yielding point for ductile materials such as Aluminum alloy, we define the emergent yielding point by the intersection of the $\varepsilon_{11}-\varepsilon^e_{11}$ curve and the straight line determined by the strain-free $\lambda_{11}$ right shifted by $2\%$ on the $\varepsilon^e_{11}$-axis. And we define the failure point to be the point where a structural phase transition occurs. $\varepsilon_Y$ provides the upper boundary concerning the applicability of the theory of emergent elasticity\cite{hu2016emergent}, while $\varepsilon_F$ denotes the critical strain where a drastic change of properties of SkX is about to occur. From FIG. 1(b, c), we learn that $\varepsilon_Y$ decreases with $b$ while $\varepsilon_F$ increases with $b$, both of which are induced by the decrease of strain-free $\lambda_{11}$ with $b$. To be more specific, consider the $\varepsilon_{11}-\varepsilon^e_{11}$ curve at $b=0.3$, $\varepsilon^e_{11}$ first decreases and then increases with $\varepsilon_{11}$, indicating a drastic change of $\lambda_{11}$ with $\varepsilon_{11}$, and hence $\varepsilon_Y$ is small since it describes the region where $\lambda_{11}$ approximately keeps its initial strain-free value. Meanwhile, $\varepsilon_F$ is large because larger value of $\varepsilon_{11}$ is required to achieve the same level of $\varepsilon^e_{11}$ than the case where $b$ is smaller.
\begin{figure*}
    \centering
    \includegraphics[scale=0.6]{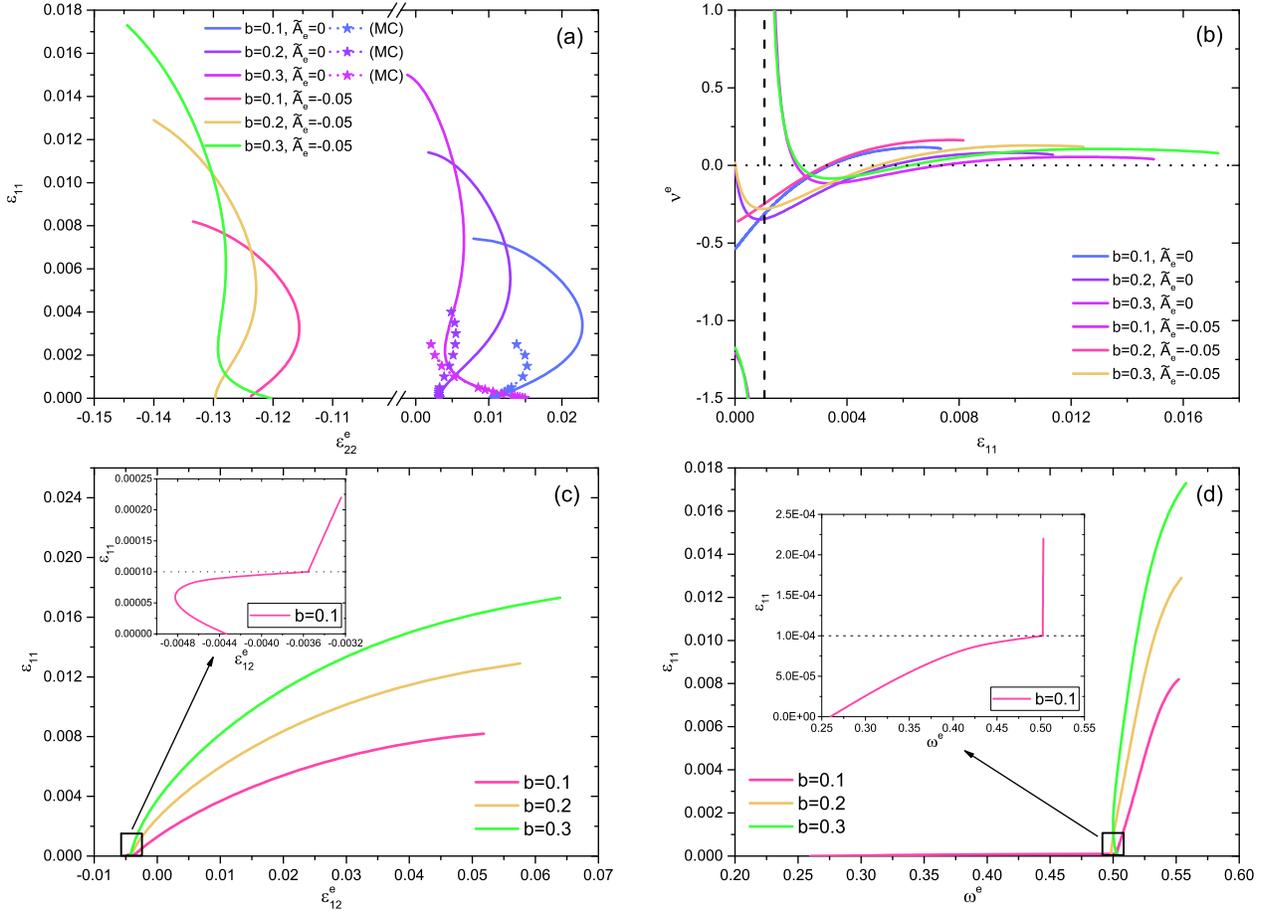}
    \caption{Variation of $(a)$, $(b)$ the emergent Poisson¡¯s ratio $\nu^e$, $(c)$ and $(d)$ $\omega^e$ with $\varepsilon_{11}$ for SkX in MnSi calculated at different values of rescaled magnetic field $b$ and rescaled exchange anisotropy coefficient $\widetilde A_e$. In $(c)$ and $(d)$, all the curves are plotted at the condition $\widetilde A_e=-0.05$}
    \label{f2}
\end{figure*}

FIG. 2 shows the variation of $\varepsilon^e_{22}$, $\varepsilon^e_{12}$ and $\omega^e$ with $\varepsilon_{11}$ in the range $0\leq\varepsilon_{11}\leq 0.02$, where the $\varepsilon_{11}-\varepsilon^e_{12}$ curves and $\varepsilon_{11}-\omega^e$ curves plotted in FIG. 2(c, d) are calculated at the condition $\widetilde A_e=-0.05$ (neglecting exchange anisotropy, $\varepsilon^e_{12}$ and $\omega^e$ are both zero for any $\varepsilon_{11}$). It is shown that $\varepsilon^e_{22}$ varies in a complicated way with $\varepsilon_{11}$ (FIG. 2(a)). We define the emergent Poisson¡¯s ration as $\nu^e=-\varepsilon^e_{22}/\varepsilon^e_{11}$, and plot in FIG. 2(b) the $\nu^e-\varepsilon_{11}$ curves calculated at various conditions. It is found that for all calculated conditions, SkX always possess negative emergent Poisson¡¯s ratio in a considerable range of positive $\varepsilon_{11}$. A negative emergent Poisson¡¯s ratio indicates the number of skyrmions in a unit volume of the material suffering mechanical forces is not conservative.
\begin{figure*}
    \centering
    \includegraphics[scale=0.5]{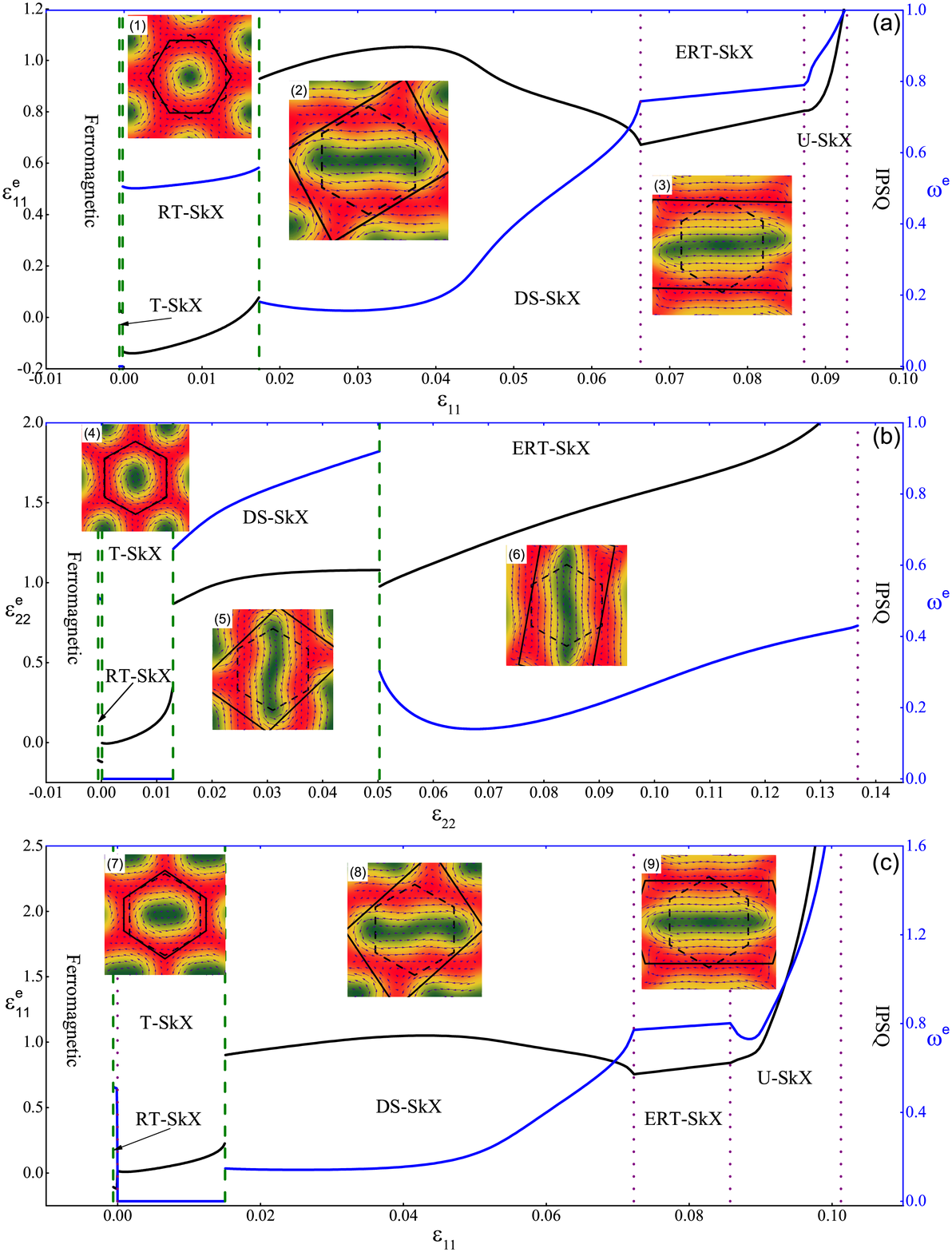}
    \caption{Structural phase transitions of the SkX phase in MnSi suffering uniaxial distortion obtained at $t=0.5,b=0.3$. $(a)$ Appearance of different magnetic states as $(a, c)$ $\varepsilon_{11}$ and $(b)$ $\varepsilon_{22}$ increases from $-0.01$, where $(a, b)$ are calculated at the condition  $\widetilde A_e=-0.05$, and (c) is calculated at the condition $\widetilde A_e=0$. The existence of different phases is characterized by two order parameters $\varepsilon^e_{11}$ (to the left axis) and $\omega^e$ (to the right axis). The insets show the typical field configuration of magnetization for the corresponding phases, where the vectors illustrate the distribution of the in-plane magnetization components with length proportional to their magnitude, while the colored density plot illustrates the distribution of the out-of-plane magnetization component.}
    \label{f3}
\end{figure*}
\newline
\newline
\noindent{\textbf{Structural transitions of skyrmion crystal in MnSi suffering uniaxial tension}}

The break points appearing near $\varepsilon_{11}=0.0001$ in the $\varepsilon_{11}-\varepsilon^e_{11}$, $\varepsilon_{11}-\varepsilon^e_{12}$ and $\varepsilon_{11}-\omega^e$ curves at $b=0.1$ shown in FIG. 1, 2 imply the occurrence of a structural phase transition of the SkX, accompanied by emergent rotation (since $\omega^e$ changes sharply). By studying the metastability and phase transitions of SkX in MnSi at $t=0.5$, $b=0.3$ suffering uniaxial distortion, we find that $\mathbf a)$ 6 types of nontrivial emergent crystalline states may appear in MnSi suffering uniaxial tension, distinguished by a difference of emergent elastic property. They are: triangle skyrmion crystal (T-SkX), rotated triangle skyrmion crystal (RT-SkX), deformed square skyrmion crystal (DS-SkX), elongated and rotated triangle skyrmion crystal (ERT-SkX), unstable skyrmion crystal (U-SkX), and in-plane single-Q (IPSQ) phase. The field configuration of these different phases are shown in the insets of FIG. 3. $\mathbf b$) For small positive $\varepsilon_{11}$, T-SkX and RT-SkX are the two possible states that may appear, where RT-SkX is obtained by rotating the T-SkX by about $30^\circ$. Neglecting exchange anisotropy, T-SkX is the only stable state in this region, which switches to RT-SkX when $\widetilde A_e=-0.05$. If we change the sign of $\varepsilon_{11}$ to apply a small compression, two structural transitions will swiftly occur in turn, first from T-SkX to RT-SkX (or vice versa) and then to the ferromagnetic phase. $\mathbf c$) As $\varepsilon_{11}$ further increases, a triangle-square transition of SkX will always occur leading to appearance of the DS-SkX phase. The local field configuration of the emergent ¡°particle¡± in the DS-SkX phase is always notably elongated in the direction of uniaxial distortion (FIG. 3(a, b)). Interestingly, we find that immediately after the phase transition, the change of emergent elastic strain $\Delta\varepsilon^e_{11}$ compared with the strain-free SkX is about 60 times larger than the uniaxial strain $\varepsilon_{11}$ applied to the underlying material. This result agrees quantitatively well with the significant deformation of the SkX observed in FeGe thin film suffering uniaxial distortion\cite{shibata2015large}. Concerning the similarity of crystalline structure between MnSi and FeGe, we argue that this phase transition from RT-SkX to DS-SkX explains the exotic emergent elasticity of SkX in FeGe. $\mathbf d)$ As $\varepsilon_{11}$ further increases, the DS-SkX will transform to the ERT-SkX phase, where the emergent ¡°particle¡± is severely elongated in the direction of uniaxial distortion, and the lattice orientation and direction of elongation is are dominated by the loading direction. Neglecting exchange anisotropy (FIG. 3(c)), the ¡°long edges¡± of the elongated skyrmions in the ERT-SkX phase are exactly parallel to the direction of uniaxial distortion, while for $\widetilde A_e=-0.05$ (FIG. 3(a, b)), these ¡°long edges¡± are slightly rotated. We also notice that the location of phase transition point from DS-SkX to ERT-SkX is sensitive to the loading direction (FIG. 3(a, b)). $\mathbf {e)}$ As $\varepsilon_{11}$ further increases, a phase transition from the ERT-SkX phase to the U-SkX phase may occur, depending on the loading direction. When the uniaxial distortion is applied in the x-axis (FIG. 3(a, c)), we see a clear sign of second order phase transition by observing the variation of order parameters with $\varepsilon_{11}$, indicating appearance of the U-SkX phase, which is mechanically very ¡°soft¡± and unstable. It collapses into the IPSQ phase easily under further extension. On the other hand, if we apply uniaxial distortion in the y-axis (FIG. 3(b)), ERT-SkX is the only stable state in the same range of $\varepsilon_{11}$. As $\varepsilon_{11}$ increases, the ERT-SkX phase also becomes gradually ¡°softer¡± and finally transform to the IPSQ phase. $\mathbf{f)}$ In FIG. 3, we choose $\varepsilon^e_{11}$ and $\omega^e$ as the order parameters of the system, and plot their variation with $\varepsilon_{11}$ to illustrate the behavior of different phases. A novel behavior of SkX as well as any emergent crystalline states is the general existence and variation of $\omega^e$ when subject to mechanical forces. Emergent rotation exists for two major reasons. Firstly, as $\varepsilon_{11}$ increases, the anisotropy of the free energy density functional is gradually changing, which renders a change of lattice direction of SkX to achieve lower free energy density. Secondly, the structures of different emergent crystalline states (e.g., T-SkX and DS-SkX) appearing on the two sides of a transition are usually directionally incompatible with each other. Therefore, emergent rotation is inevitable which leads to jump of $\omega^e$ during these phase transitions. Emergent rotation with respect to the underlying atomic lattices is an important and unique way for SkX and other emergent crystals to resist tension of the material. Nevertheless, since emergent rotation breaks the ordinary periodic boundary condition widely used in various calculations, it is hard to take into account when using pure numerical simulation methods\cite{wang2018uniaxial}.

Deformation of emergent cyrstals is essentially different from that of ordinary crystals: consider SkX as an example, its deformation is realized microscopically by variation of magnetization distribution, while a variation of the distance between two neighboring spins is not necessary. Based on this fundamental difference, emergent crystals possess two unique features compared with ordinary crystals: $\mathbf {a)}$ the local field configuration of the emergent particles inside an emergent crystal is geometrically unrestricted. The emergent particles in an emergent crystal are composed of multiple atoms or their electric and magnetic degrees of freedom with a nonconservative number. I.e., the emergent deformation is realized by ¡°absorbing¡± the field configuration of neighboring atoms inside the emergent particle, for which its magnitude of deformability is unlimited. $\mathbf {b)}$ The emergent crystals can undergo emergent rotation with respect to the underlying material. We find that these two features dominates the nonelastic behavior of SkX suffering all kinds of external fields, including not only mechanical loading, but also magnetic field\cite{wang2017enhanced}, electric field\cite{jiang2017direct}, and temperature field\cite{morikawa2017deformation}. Moreover, these features permit a wide variety of possible crystalline structures of emergent crystals suffering different anisotropic fields, as shown in the results obtained here for SkX.
\begin{figure}
    \centering
    \includegraphics[scale=0.28]{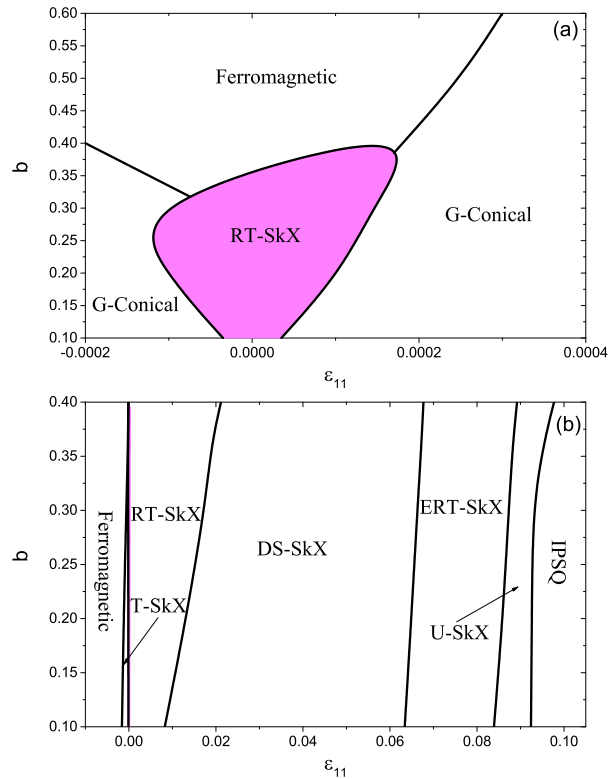}
    \caption{$\varepsilon_{11}-b$ phase diagram of $(a)$ equilibrium magnetic states of MnSi and $(b)$ metastabiliy of SkX in MnSi calculated at the condition $\widetilde A_e=0$.}
    \label{f4}
\end{figure}
\newline
\newline
\noindent{\textbf{Strain-magnetic field Phase diagrams for MnSi}}

 We further study the phase diagrams for MnSi suffering uniaxial distortion. We plot two types of $\varepsilon_{11}-b$ phase diagrams at the condition $t=0.5$, $\widetilde A_e=-0.05$: $\varepsilon_{11}-b$ phase diagram of equilibrium states (FIG. 4(a)), and $\varepsilon_{11}-b$ phase diagram of metastability for SkX  (FIG. 4(b)). It is found that thermodynamically stable SkX exists in a very narrow range of $\varepsilon_{11}$, and exists in only one state: the RT-SkX phase. On the other hand, concerning the robustness of SkX\cite{karube2016robust,hu2018effect}, metastable SkX exists in a much wider range of $\varepsilon_{11}$ and presence in a variety of different phases.

Concerning the wave nature of emergent crystals\cite{hu2018effect}, all metastable states are stabilized by nonlinear mode-mode interaction, while the corresponding thermodynamically stable state: the generalized-conical (G-conical) phase is a single-Q structure whose wave vector is free to rotate in space. That is to say, the phase transition from any metastable emergent crystalline state to the G-conical is realized by cancelling of all the nonlinear mode-mode interactions (i.e., to set all related Fourier amplitudes to zero). This is physically difficult, because unlike annihilating a localized skyrmion, any Fourier component of magnetization exist globally and is hard to cancel through any local fluctuation. As a result, besides the well known topological protection attributed to isolated skyrmions, SkX is additionally protected globally by mode-mode interactions. In this sense, once we have a stable strain-free SkX phase, the appearance of various types of metastable SkX predicted in our calculation is anticipated, especially at low temperature range. This viewpoint is supported by the successful observation of the DS-SkX phase in uniaxially stretched FeGe\cite{shibata2015large}.
\\
\\
\noindent{\textbf{Methods}}

$Emergent\, elasticity\, for\, 2D\, hexagonal\, emergent\, \\crystals\, in\, helimagnets.$

To formulate the theory of emergent elasticity\cite{hu2016emergent} for emergent crystals in helimagnets, we switch from eq. (\ref{1}) to a more generalized rescaled free energy functional\cite{hu2017unified}
\begin{equation}
    \begin{aligned}
        \widetilde w(\mathbf m,\varepsilon_{ij})=&\sum^3_{i=1}\left(\frac{\partial\mathbf m}{\partial r_i}\right)^2+\mathbf m\cdot(\nabla\times\mathbf m)
        -2\mathbf b\cdot\mathbf m+\\ &t\mathbf m^2+\mathbf m^4
        +\sum^3_{i=1}\left[\widetilde A_c m^4_i+\widetilde A_e \left(\frac{\partial m_i}{\partial r_i}\right)^2\right]\\ &+\widetilde w_{el}+\widetilde w_{me},
    \end{aligned}
    \label{5}
\end{equation}
where
\begin{equation}
    \begin{aligned}
        \widetilde w_{el}=&\frac{1}{2}\widetilde C_{11}(\varepsilon^2_{11}+\varepsilon^2_{22}+\varepsilon^2_{33})+\widetilde C_{12}(\varepsilon^2_{11}\varepsilon^2_{22}+\varepsilon^2_{22}\varepsilon^2_{33}\\ &+\varepsilon^2_{22}\varepsilon^2_{33})+\frac{1}{2}\widetilde C_{44}(\gamma^2_{12}+\gamma^2_{13}+\gamma^2_{23})
    \end{aligned}
    \label{6}
\end{equation}
denotes the elastic energy density of materials with cubic symmetry, $\gamma_{ij}=2\varepsilon_{ij}$ denote the engineering shear strains, and
\begin{equation}
    \begin{aligned}
        \widetilde w_{me}=&\widetilde K\mathbf m^2\varepsilon_{ii}+\widetilde L_1(m^2_1\varepsilon_{11}+m^2_2\varepsilon_{22}+m^2_3\varepsilon_{33})
        \\ &+\widetilde L_2(m^2_3\varepsilon_{11}+m^2_1\varepsilon_{22}+m^2_2\varepsilon_{33})+\widetilde L_3(m_1m_2\gamma_{12}\\ &+m_1m_3\gamma_{13}+m_2m_3\gamma_{23})+\sum^6_{i=1}\widetilde L_{Oi}\widetilde f_{Oi},
    \end{aligned}
    \label{7}
\end{equation}
denotes the magnetoelastic energy density, where
\begin{equation}
    \begin{aligned}
        \widetilde f_{O1}=&\varepsilon_{11}(m_{1,2}m_3-m_{1,3}m_2)+\varepsilon_{22}(m_{2,3}m_1\\ &-m_{2,1}m_3)+\varepsilon_{33}(m_{3,1}m_2-m_{3,2}m_1),
        \\\widetilde f_{O2}=&\varepsilon_{11}(m_{3,2}m_2-m_{2,1}m_3)+\varepsilon_{22}(m_{1,2}m_3\\ &-m_{3,2}m_1)+\varepsilon_{33}(m_{2,3}m_1-m_{1,3}m_2),
        \\\widetilde f_{O3}=&\varepsilon_{11}m_1(m_{2,3}-m_{3,2})+\varepsilon_{22}m_2(m_{3,1}-m_{1,3})\\ &+\varepsilon_{33}m_3(m_{1,2}-m_{2,1}),
        \\\widetilde f_{O4}=&\gamma_{23}(m_{1,3}m_3-m_{1,2}m_2)+\gamma_{13}(m_{2,1}m_1\\ &-m_{2,3}m_3)+\gamma_{12}(m_{3,2}m_2-m_{3,1}m_1),
        \\\widetilde f_{O4}=&\gamma_{23}(m_{1,3}m_3-m_{1,2}m_2)+\gamma_{13}(m_{2,1}m_1\\ &-m_{2,3}m_3)+\gamma_{12}(m_{3,2}m_2-m_{3,1}m_1),
        \\\widetilde f_{O5}=&\gamma_{23}(m_{3,1}m_3-m_{2,1}m_2)+\gamma_{13}(m_{1,2}m_1\\ &-m_{3,2}m_3)+\gamma_{12}(m_{2,3}m_2-m_{1,3}m_1),
        \\\widetilde f_{O6}=&\gamma_{23}m_1(m_{3,3}-m_{2,2})+\gamma_{13}m_2(m_{1,1}-m_{3,3})\\ &+\gamma_{12}m_3(m_{2,2}-m_{1,1}).
    \end{aligned}
    \label{8}
\end{equation}
Eq. (\ref{1}) can be derived from eq. (\ref{5}) by setting all components of elastic strains to zero except $\varepsilon_{11}$. For deformable hexagonal 2-D emergent crystals, the rescaled magnetization $\mathbf m$ can be described mathematically as a Fourier series\cite{hu2018wave}:
\begin{equation}
    \begin{aligned}
        \mathbf m=\mathbf m_0+\sum^\infty_{i=1}\sum^{n_i}_{j=1}\mathbf m_{\mathbf q_{ij}}e^{{\rm i}\mathbf q_{ij}\cdot\mathbf r}.
    \end{aligned}
    \label{9}
\end{equation}
It is convenient to expand $\mathbf m_{\mathbf q_{ij}}$ as $\mathbf m_{\mathbf q_{ij}}=\tilde c_{ij1}\mathbf P_{ij1}+\tilde c_{ij2}\mathbf P_{ij2}+\tilde c_{ij3}\mathbf P_{ij3}$, where $\tilde c_{ij1}=\tilde c^{re}_{ij1}+{\rm i}\tilde c^{im}_{ij1}$, $\tilde c_{ij2}=\tilde c^{re}_{ij2}+{\rm i}\tilde c^{im}_{ij2}$, $\tilde c_{ij3}=\tilde c^{re}_{ij3}+{\rm i}\tilde c^{im}_{ij3}$ are complex variables to be determined, and $\mathbf P_{ij1}=\frac{1}{\sqrt{2}s_iq}[-iq_{ijy}, iq_{ijx}, s_iq]^T$, $\mathbf P_{ij2}=\frac{1}{s_iq}[q_{ijx}, q_{ijy}, 0]^T$, $\mathbf P_{ij3}=\frac{1}{\sqrt{2}s_iq}[iq_{ijy}, -iq_{ijx}, s_iq]^T$ with $\mathbf q_{ij}=[q_{ijx}, q_{ijy}]^T$, $|q_{ij}|=s_iq$. Assume that the undeformed wave vectors to be $q_{11}^0=q[0,1]^T$, $q_{12}^0=q[-\frac{\sqrt{3}}{2},-\frac{1}{2}]^T$ for a hexagonal lattice, the deformed wave vectors are related to the emergent elastic strains and emergent rotation angle by\cite{hu2016emergent}
\begin{equation}
    \begin{aligned}
        \mathbf q_{11}=&\frac{q}{s}[\omega^e-\varepsilon^e_{12},1+\varepsilon^e_{11}]^T,
        \\\mathbf q_{12}=&\frac{q}{2s}[-\sqrt{3}-\sqrt{3}\varepsilon^e_{22}+\varepsilon^e_{12}-\omega^e,\\ &-1-\varepsilon^e_{11}+\sqrt{3}(\varepsilon^e_{12}+\omega^e)]^T,
    \end{aligned}
    \label{10}
\end{equation}
where $s=1+\varepsilon^e_{11}+\varepsilon^e_{22}+\varepsilon^e_{11}\varepsilon^e_{22}-(\varepsilon^e_{12})^2+(\omega^e)^2$. The averaged rescaled free energy density for emergent crystalline states can be obtained by substituting eqs. (\ref{9}, \ref{10}) into eq. (\ref{5}) and taking the volume average $\bar{w}=\frac{1}{V}\int \widetilde w dV$. At small disturbance around a metastable state and neglecting possible periodic elastic fields, the averaged rescaled free energy density for emergent crystalline states can be expanded in a quadratic form of all the independent variables as
\begin{equation}
    \begin{aligned}
        \bar w=&\bar w_u+\frac{1}{2}(d\bm{\varepsilon}^{e})^T\widetilde{\mathbf C}^{ea} d\bm{\varepsilon}^{ea}+\frac{1}{2}(d\mathbf m^a)^T\widetilde{\bm\mu}^a d\mathbf m^a
        \\ &+\frac{1}{2}(d\bm{\varepsilon})^T\widetilde{\mathbf C} d\bm{\varepsilon}+(d\bm{\varepsilon})^T\widetilde{\mathbf h} d\bm{\varepsilon}^{ea}
        \\ &+(d\bm{\varepsilon}^{ea})^T\widetilde{\mathbf g}^{em} d\mathbf{m}^{a}+(d\bm{\varepsilon})^T\widetilde{\mathbf g}^m d\mathbf{m}^{a},
    \end{aligned}
    \label{11}
\end{equation}
where $\bar w_u$ denotes the undisturbed averaged rescaled Helmholtz free energy density, and
\begin{equation}
    \begin{aligned}
        \bm{\varepsilon}^{ea}=[\varepsilon^e_{11},\varepsilon^e_{22},\varepsilon^e_{12},\bm{\omega}^e]^T,
    \end{aligned}
    \label{12}
\end{equation}
\begin{equation}
    \begin{aligned}
        \bm{\varepsilon}=[\varepsilon_{11},\varepsilon_{22},\varepsilon_{12},\gamma_{23},\gamma_{13},\gamma_{12}]^T,
    \end{aligned}
    \label{13}
\end{equation}
\begin{equation}
    \begin{aligned}
        \mathbf m^a=[&m_{01},m_{02},m_{03},\tilde c^{re}_{111},\tilde c^{im}_{112},\tilde c^{re}_{113},\tilde c^{re}_{121},
        \\ &\tilde c^{im}_{122},\tilde c^{re}_{123},\tilde c^{re}_{131},\tilde c^{im}_{132},\tilde c^{re}_{133},\tilde c^{re}_{211},\cdot\cdot\cdot]^T.
    \end{aligned}
    \label{14}
\end{equation}
\begin{equation}
    \begin{aligned}
        & \widetilde C_{ij}=\left(\frac{\partial^2 \bar{w}}{\partial\varepsilon_i\partial\varepsilon_j}\right)_0,
        \widetilde h_{ij}=\left(\frac{\partial^2 \bar{w}}{\partial\varepsilon_i\partial\varepsilon^{ea}_j}\right)_0,
        \\ & \widetilde g^m_{ij}=\left(\frac{\partial^2 \bar{w}}{\partial\varepsilon_i\partial m^a_j}\right)_0,
        \widetilde C^e_{ij}=\left(\frac{\partial^2 \bar{w}}{\partial\varepsilon^{ea}_i\partial\varepsilon^{ea}_j}\right)_0,
        \\ &\widetilde g^{em}_{ij}=\left(\frac{\partial^2 \bar{w}}{\partial\varepsilon^{ea}_i\partial m^a_j}\right)_0,
        \widetilde \mu^a_{ij}=\left(\frac{\partial^2 \bar{w}}{\partial m^a_i\partial m^a_j}\right)_0.
    \end{aligned}
    \label{15}
\end{equation}
One should notice that the form of eq. (\ref{14}) has been simplified after canceling the degrees of freedom related to the in-plane rigid translation\cite{hu2016emergent}.
The linear constitutive equations are derived from eq. (\ref{11}) as:
\begin{equation}
    \begin{aligned}
        &d\bm{\sigma}^{ea}=\widetilde{\mathbf C}^e d\bm{\varepsilon}^{ea}+\widetilde{\mathbf h}^T d\bm{\varepsilon}+\widetilde{\mathbf g}^{em}d\mathbf m^a,
        \\ &d\bm{\sigma}=\widetilde{\mathbf C} d\bm{\varepsilon}^{ea}+\widetilde{\mathbf h} d\bm{\varepsilon}^{ea}+\widetilde{\mathbf g}^{m}d\mathbf m^a,
        \\ & d\mathbf B^a=\bm{\mu}^a d\mathbf{m}^a+(\widetilde{\mathbf g}^m)^T d\bm{\varepsilon}+(\widetilde{\mathbf g}^{em})^T d\bm{\varepsilon}^{ea},
    \end{aligned}
    \label{16}
\end{equation}
where
\begin{equation}
    \begin{aligned}
        \bm{\sigma}^{ea}=[\sigma^e_{11},\sigma^e_{22},\sigma^e_{12},\Gamma^e_{12}]^T ,
    \end{aligned}
    \label{17}
\end{equation}
\begin{equation}
    \begin{aligned}
        \bm{\sigma}=[\sigma_{11},\sigma_{22},\sigma_{33},\sigma_{23},\sigma_{13},\sigma_{12}]^T ,
    \end{aligned}
    \label{18}
\end{equation}
\begin{equation}
    \begin{aligned}
        \mathbf B^a=[&B_1,B_2,B_3,\tilde d^{re}_{111},\tilde d^{im}_{112},\tilde d^{re}_{113},\tilde d^{re}_{121},
        \\ &\tilde d^{im}_{122},\tilde d^{re}_{123},\tilde d^{re}_{131},\tilde d^{im}_{132},\tilde d^{re}_{133},\tilde d^{re}_{211},\cdot\cdot\cdot]^T
    \end{aligned}
    \label{19}
\end{equation}
denote work conjugates of $\bm{\varepsilon}^{ea}$, $\bm{\varepsilon}$, and $\mathbf m^a$, respectively.
When the material is subject to a disturbance of elastic strains, we have $d\mathbf B^a=\mathbf 0$ and $d\bm{\sigma}^{ea}=\mathbf 0$. In this case, after deduction we have from eq. (\ref{16})
\begin{equation}
    \begin{aligned}
        d\bm{\varepsilon}^{ea}=\bm{\lambda}d\bm{\varepsilon},
    \end{aligned}
    \label{20}
\end{equation}
where
\begin{equation}
    \begin{aligned}
        &\bm{\lambda}=-(\widetilde{\mathbf C}^{e*})^{-1}(\widetilde{\mathbf h}^*)^T,
        \\ &\widetilde{\mathbf C}^{e*}=\widetilde{\mathbf C}^e-\widetilde{\mathbf g}^{em}(\widetilde{\bm \mu}^a)^{-1}(\widetilde{\mathbf g}^{em})^T,
        \\ &\widetilde{\mathbf h}^{e*}=\widetilde{\mathbf h}^e-\widetilde{\mathbf g}^{m}(\widetilde{\bm \mu}^a)^{-1}(\widetilde{\mathbf g}^{em})^T.
    \end{aligned}
    \label{21}
\end{equation}
Eq. (\ref{20}) provides the linear relation between the emergent elastic strains, emergent rotations and the elastic strains, and $\bm{\lambda}$ is called the crossover strain ratio (CSR) matrix.

One should notice that the compliance matrices defined in eq. (\ref{15}) provide necessary information to determine the metastability of a magnetic phase. For a state obtained from free energy minimization based on Fourier representation, the positive definiteness of the following matrix\cite{hu2018wave} guarantees the intrinsic stability of the state:
\begin{equation}
    \bm \Phi=
    \begin{bmatrix}
        \mathbf {\widetilde C}^e & \mathbf {\widetilde g}^{em}
        \\ (\mathbf {\widetilde g}^{em})^T & \bm {\widetilde \mu}^a
    \end{bmatrix}.
    \label{22}
\end{equation}
When the positive definiteness of $\bm\Phi$ is destroyed, the eigenvector corresponding to the smallest eigenvalue (softest mode) of $\bm\Phi$ indicates the direction of evolution for the system from the current state.
\\
\\
$Free\, energy\, minimization\, based\, on\, Monte-Carlo\\ Simulation. $

We perform Monte-Carlo Simulation based on annealing algorithm to minimize the free energy of the system. To process, we discretize the free energy density functional shown in eq. (\ref{1}) with 2-D grids in Cartesian coordinates. Periodic boundary conditions are used in the calculation. For given elastic strains, the energy density functional can be written as
\begin{equation}
    \begin{aligned}
        \widetilde w =\widetilde w(m_x(x,y),m_y(x,y),m_z(x,y)),
    \end{aligned}
    \label{23}
\end{equation}
where $m_x$, $m_y$, $m_z$ denote the components of the rescaled magnetization $\mathbf m$.

We use the central difference method to calculate the value of partial derivatives
\begin{equation}
    \begin{aligned}
        \frac{\partial m_k}{\partial x}[i][j]&=\frac{m_k[i+1][j]-m_k[i-1][j]}{2\Delta x},
        \\ \frac{\partial m_k}{\partial y}[i][j]&=\frac{m_k[i+1][j]-m_k[i-1][j]}{2\Delta y},
        \\ k&=x,y,z,
    \end{aligned}
    \label{24}
\end{equation}
where the values of $\Delta x$ and $\Delta y$ determine the error of the Monte Carlo simulation.

In order to calculate the emergent strain of SkX, $\Delta x$ and $\Delta y$ are also set as variables. In this case, we calculated the averaged free energy density for different values of $\Delta x$ and $\Delta y$ to obtain the smallest one. And corresponding values of $\Delta x$ and $\Delta y$ determine the emergent elastic strains of SkX
\begin{equation}
    \begin{aligned}
        &\varepsilon^e_{11}=\frac{\Delta x^\prime-\Delta x}{\Delta x}
        \\ &\varepsilon^e_{22}=\frac{\Delta y^\prime-\Delta y}{\Delta y}
    \end{aligned}
    \label{25}
\end{equation}
This method is only applicable to moderate emergent deformation of SkX. When the difference between equilibrium values of $\Delta x$ and $\Delta y$ is too large, the precision can no longer be guaranteed, and a new set of grids is needed.
\\
\\
\\
\\
\begin{acknowledgments}
Y.H. conceived the idea and conducted the calculation based on Fourier representation. X.L. conducted the calculation based on Monte Carlo simulation. Y.H. and X.L. co-wrote the paper. Y.H., X.L. and B.W. discussed the results for revision. The work was supported by the NSFC (National Natural Science Foundation of China) through the funds 11772360, 11832019, 11472313, 11572355 and Pearl River Nova Program of Guangzhou (Grant No. 201806010134).
\end{acknowledgments}

\bibliographystyle{nature}
\bibliography{Manuscript}

\begin{thebibliography}{10}

\bibitem{muhlbauer2009skyrmion}
M{\"u}hlbauer, S., Binz, B., Jonietz, F., Pfleiderer, C., Rosch, A., Neubauer,
  A., Georgii, R., and B{\"o}ni, P.
\newblock {\em Science}{ \bf 323}(5916), 915--919 (2009).

\bibitem{kezsmarki2015neel}
K{\'e}zsm{\'a}rki, I., Bord{\'a}cs, S., Milde, P., Neuber, E., Eng, L., White,
  J., R{\o}nnow, H.~M., Dewhurst, C., Mochizuki, M., Yanai, K., et~al.
\newblock {\em Nature Materials}{ \bf 14}(11), 1116 (2015).

\bibitem{von2015influence}
von Bergmann, K., Menzel, M., Kubetzka, A., and Wiesendanger, R.
\newblock {\em Nano Letters}{ \bf 15}(5), 3280--3285 (2015).

\bibitem{langner2014coupled}
Langner, M., Roy, S., Mishra, S., Lee, J., Shi, X., Hossain, M., Chuang, Y.-D.,
  Seki, S., Tokura, Y., Kevan, S., et~al.
\newblock {\em Physical Review Letters}{ \bf 112}(16), 167202 (2014).

\bibitem{tanigaki2015real}
Tanigaki, T., Shibata, K., Kanazawa, N., Yu, X., Onose, Y., Park, H.~S.,
  Shindo, D., and Tokura, Y.
\newblock {\em Nano Letters}{ \bf 15}(8), 5438--5442 (2015).

\bibitem{lee2015zero}
Lee, S.-Y. and Han, J.~H.
\newblock {\em Physical Review B}{ \bf 91}(24), 245121 (2015).

\bibitem{huang2017stabilization}
Huang, S., Zhou, C., Chen, G., Shen, H., Schmid, A.~K., Liu, K., and Wu, Y.
\newblock {\em Physical Review B}{ \bf 96}(14), 144412 (2017).

\bibitem{ackerman2015self}
Ackerman, P.~J., Van De~Lagemaat, J., and Smalyukh, I.~I.
\newblock {\em Nature Communications}{ \bf 6}, 6012 (2015).

\bibitem{bauerle1996laboratory}
B{\"a}uerle, C., Bunkov, Y.~M., Fisher, S., Godfrin, H., and Pickett, G.
\newblock {\em Nature}{ \bf 382}(6589), 332 (1996).

\bibitem{al2001skyrmions}
Al~Khawaja, U. and Stoof, H.
\newblock {\em Nature}{ \bf 411}(6840), 918 (2001).

\bibitem{rossler2006spontaneous}
R{\"o}{\ss}ler, U., Bogdanov, A., and Pfleiderer, C.
\newblock {\em Nature}{ \bf 442}(7104), 797 (2006).

\bibitem{fu2016persistent}
Fu, J., Penteado, P.~H., Hachiya, M.~O., Loss, D., and Egues, J.~C.
\newblock {\em Physical Review Letters}{ \bf 117}(22), 226401 (2016).

\bibitem{nych2017spontaneous}
Nych, A., Fukuda, J.-i., Ognysta, U., {\v{Z}}umer, S., and Mu{\v{s}}evi{\v{c}},
  I.
\newblock {\em Nature Physics}{ \bf 13}(12), 1215 (2017).

\bibitem{karube2016robust}
Karube, K., White, J., Reynolds, N., Gavilano, J., Oike, H., Kikkawa, A.,
  Kagawa, F., Tokunaga, Y., R{\o}nnow, H.~M., Tokura, Y., et~al.
\newblock {\em Nature Materials}{ \bf 15}(12), 1237 (2016).

\bibitem{johnson2013mnsb}
Johnson, R., Cao, K., Chapon, L., Fabrizi, F., Perks, N., Manuel, P., Yang, J.,
  Oh, Y.~S., Cheong, S.-W., and Radaelli, P.
\newblock {\em Physical Review Letters}{ \bf 111}(1), 017202 (2013).

\bibitem{leonov2015multiply}
Leonov, A. and Mostovoy, M.
\newblock {\em Nature Communications}{ \bf 6}, 8275 (2015).

\bibitem{yu2010real}
Yu, X., Onose, Y., Kanazawa, N., Park, J., Han, J., Matsui, Y., Nagaosa, N.,
  and Tokura, Y.
\newblock {\em Nature}{ \bf 465}(7300), 901 (2010).

\bibitem{heinze2011spontaneous}
Heinze, S., Von~Bergmann, K., Menzel, M., Brede, J., Kubetzka, A.,
  Wiesendanger, R., Bihlmayer, G., and Bl{\"u}gel, S.
\newblock {\em Nature Physics}{ \bf 7}(9), 713 (2011).

\bibitem{yu2011near}
Yu, X., Kanazawa, N., Onose, Y., Kimoto, K., Zhang, W., Ishiwata, S., Matsui,
  Y., and Tokura, Y.
\newblock {\em Nature Materials}{ \bf 10}(2), 106 (2011).

\bibitem{jiang2017skyrmions}
Jiang, W., Chen, G., Liu, K., Zang, J., te~Velthuis, S.~G., and Hoffmann, A.
\newblock {\em Physics Reports}{ \bf 704}, 1--49 (2017).

\bibitem{du2015electrical}
Du, H., Liang, D., Jin, C., Kong, L., Stolt, M.~J., Ning, W., Yang, J., Xing,
  Y., Wang, J., Che, R., et~al.
\newblock {\em Nature Communications}{ \bf 6}, 7637 (2015).

\bibitem{ambrose2013melting}
Ambrose, M. and Stamps, R.
\newblock {\em New Journal of Physics}{ \bf 15}(5), 053003 (2013).

\bibitem{jonietz2010spin}
Jonietz, F., M{\"u}hlbauer, S., Pfleiderer, C., Neubauer, A., M{\"u}nzer, W.,
  Bauer, A., Adams, T., Georgii, R., B{\"o}ni, P., Duine, R.~A., et~al.
\newblock {\em Science}{ \bf 330}(6011), 1648--1651 (2010).

\bibitem{zang2011dynamics}
Zang, J., Mostovoy, M., Han, J.~H., and Nagaosa, N.
\newblock {\em Physical Review Letters}{ \bf 107}(13), 136804 (2011).

\bibitem{neubauer2009topological}
Neubauer, A., Pfleiderer, C., Binz, B., Rosch, A., Ritz, R., Niklowitz, P., and
  B{\"o}ni, P.
\newblock {\em Physical Review Letters}{ \bf 102}(18), 186602 (2009).

\bibitem{litzius2017skyrmion}
Litzius, K., Lemesh, I., Kr{\"u}ger, B., Bassirian, P., Caretta, L., Richter,
  K., B{\"u}ttner, F., Sato, K., Tretiakov, O.~A., F{\"o}rster, J., et~al.
\newblock {\em Nature Physics}{ \bf 13}(2), 170 (2017).

\bibitem{jiang2017direct}
Jiang, W., Zhang, X., Yu, G., Zhang, W., Wang, X., Jungfleisch, M.~B., Pearson,
  J.~E., Cheng, X., Heinonen, O., Wang, K.~L., et~al.
\newblock {\em Nature Physics}{ \bf 13}(2), 162 (2017).

\bibitem{shiomi2013topological}
Shiomi, Y., Kanazawa, N., Shibata, K., Onose, Y., and Tokura, Y.
\newblock {\em Physical Review B}{ \bf 88}(6), 064409 (2013).

\bibitem{white2014electric}
White, J., Pr{\v{s}}a, K., Huang, P., Omrani, A., {\v{Z}}ivkovi{\'c}, I.,
  Bartkowiak, M., Berger, H., Magrez, A., Gavilano, J., Nagy, G., et~al.
\newblock {\em Physical Review Letters}{ \bf 113}(10), 107203 (2014).

\bibitem{shibata2015large}
Shibata, K., Iwasaki, J., Kanazawa, N., Aizawa, S., Tanigaki, T., Shirai, M.,
  Nakajima, T., Kubota, M., Kawasaki, M., Park, H., et~al.
\newblock {\em Nature Nanotechnology}{ \bf 10}(7), 589--592 (2015).

\bibitem{okamura2016transition}
Okamura, Y., Kagawa, F., Seki, S., and Tokura, Y.
\newblock {\em Nature Communications}{ \bf 7}, 12669 (2016).

\bibitem{nii2015uniaxial}
Nii, Y., Nakajima, T., Kikkawa, A., Yamasaki, Y., Ohishi, K., Suzuki, J.,
  Taguchi, Y., Arima, T., Tokura, Y., and Iwasa, Y.
\newblock {\em Nature Communications}{ \bf 6}, 8539 (2015).

\bibitem{chacon2015uniaxial}
Chacon, A., Bauer, A., Adams, T., Rucker, F., Brandl, G., Georgii, R., Garst,
  M., and Pfleiderer, C.
\newblock {\em Physical Review Letters}{ \bf 115}(26), 267202 (2015).

\bibitem{wang2018uniaxial}
Wang, J., Shi, Y., and Kamlah, M.
\newblock {\em Physical Review B}{ \bf 97}, 024429 Jan  (2018).

\bibitem{karhu2010structure}
Karhu, E., Kahwaji, S., Monchesky, T., Parsons, C., Robertson, M., and
  Maunders, C.
\newblock {\em Physical Review B}{ \bf 82}(18), 184417 (2010).

\bibitem{huang2012extended}
Huang, S. and Chien, C.
\newblock {\em Physical Review Letters}{ \bf 108}(26), 267201 (2012).

\bibitem{liu2017chopping}
Liu, Y., Lei, N., Zhao, W., Liu, W., Ruotolo, A., Braun, H.-B., and Zhou, Y.
\newblock {\em Applied Physics Letters}{ \bf 111}(2), 022406 (2017).

\bibitem{chen2015unlocking}
Chen, G., Kang, S.~P., Kwon, H.~Y., Won, C., Wu, Y., Qiu, Z., Schmid, A.~K.,
  et~al.
\newblock {\em Nature Communications}{ \bf 6}, 6598 (2015).

\bibitem{zhang2017ultrasonic}
Zhang, X.-X. and Nagaosa, N.
\newblock {\em New Journal of Physics}{ \bf 19}(4), 043012 (2017).

\bibitem{fobes2017versatile}
Fobes, D.~M., Luo, Y., Le{\'o}n-Brito, N., Bauer, E., Fanelli, V., Taylor, M.,
  DeBeer-Schmitt, L.~M., and Janoschek, M.
\newblock {\em Applied Physics Letters}{ \bf 110}(19), 192409 (2017).

\bibitem{kang2017elastic}
Kang, S., Kwon, H., and Won, C.
\newblock {\em Journal of Applied Physics}{ \bf 121}(20), 203902 (2017).

\bibitem{hu2016emergent}
Hu, Y. and Wang, B.
\newblock {\em arXiv preprint arXiv:1608.04840}{ \bf } (2016).

\bibitem{petrova2011spin}
Petrova, O. and Tchernyshyov, O.
\newblock {\em Physical Review B}{ \bf 84}(21), 214433 (2011).

\bibitem{hu2017unified}
Hu, Y. and Wang, B.
\newblock {\em New Journal of Physics}{ \bf 19}(12), 123002 (2017).

\bibitem{nii2014elastic}
Nii, Y., Kikkawa, A., Taguchi, Y., Tokura, Y., and Iwasa, Y.
\newblock {\em Physical Review Letters}{ \bf 113}(26), 267203 (2014).

\bibitem{li2013robust}
Li, Y., Kanazawa, N., Yu, X., Tsukazaki, A., Kawasaki, M., Ichikawa, M., Jin,
  X., Kagawa, F., and Tokura, Y.
\newblock {\em Physical Review Letters}{ \bf 110}(11), 117202 (2013).

\bibitem{hu2018effect}
Hu, Y.
\newblock {\em physica status solidi (RRL)--Rapid Research Letters}{ \bf
  12}(10), 1800247 (2018).

\bibitem{hu2018wave}
Hu, Y.
\newblock {\em Communications Physics}{ \bf 1}(1), 82 (2018).

\bibitem{wang2017enhanced}
Wang, C., Du, H., Zhao, X., Jin, C., Tian, M., Zhang, Y., and Che, R.
\newblock {\em Nano letters}{ \bf 17}(5), 2921--2927 (2017).

\bibitem{morikawa2017deformation}
Morikawa, D., Yu, X., Karube, K., Tokunaga, Y., Taguchi, Y., Arima, T.-h., and
  Tokura, Y.
\newblock {\em Nano Letters}{ \bf 17}(3), 1637--1641 (2017).

\end{thebibliography}

\end{document}